\documentclass{article}
\usepackage{spconf,amsmath,graphicx}

\usepackage{amsmath,amsfonts}
\usepackage{algorithmic}
\usepackage{algorithm}
\usepackage{array}
\usepackage[caption=false,font=normalsize,captionskip=-2pt]{subfig}
\usepackage{textcomp}
\usepackage{stfloats}
\usepackage[hidelinks]{hyperref}
\usepackage{url}
\usepackage{verbatim}
\usepackage{graphicx}
\usepackage{paralist}
\usepackage{xcolor}

\usepackage{cleveref}

\usepackage{jabbrv}

\usepackage{bm}
\usepackage{siunitx}
\usepackage{lipsum}
\usepackage{mathtools}

\usepackage{tabularx}
\usepackage{booktabs}

\usepackage{microtype}

\usepackage{placeins}

\usepackage{footmisc}

\usepackage{stmaryrd}

\usepackage[sort&compress,numbers,square]{natbib}

\ninept
\title{Quantifying Spatial Audio Quality Impairment}
\name{Karn N. Watcharasupat\thanks{Part of this work was done while K. N. Watcharasupat was supported by the American Association of University Women (AAUW) International Fellowship and the IEEE Signal Processing Society Scholarship Program.} and Alexander Lerch}
\address{Music Informatics Group, Georgia Institute of Technology, Atlanta, GA, USA\\
\{kwatcharasupat, alexander.lerch\}@gatech.edu}

\everypar{\looseness=-1 }

\def\lastaccessed{21 Nov 2023}
\begin{document}
\ninept
\maketitle
\begin{abstract}
Spatial audio quality is a highly multifaceted concept, with many interactions between environmental, geometrical, anatomical, psychological, and contextual considerations.
Methods for characterization or evaluation of the geometrical components of spatial audio quality, however, remain scarce, despite being perhaps the least subjective aspect of spatial audio quality to quantify. By considering interchannel time and level differences relative to a reference signal, it is possible to construct a signal model to isolate some of the spatial distortion. By using a combination of least-square optimization and heuristics, we propose a signal decomposition method to isolate the spatial error from a processed signal, in terms of interchannel gain leakages and changes in relative delays. This allows the computation of simple energy-ratio metrics, providing objective measures of spatial and non-spatial signal qualities, with minimal assumptions and no dataset dependency. Experiments demonstrate the robustness of the method against common spatial signal degradation introduced by, e.g., audio compression and music source separation.
\end{abstract}
\begin{keywords}
Multichannel signal processing, signal decomposition, spatial audio, quality evaluation 
\end{keywords}

\section{Introduction}


The development of spatial audio technology is intrinsically linked to the spatial hearing ability of human listeners. Human sound localization is commonly understood to be characterizable by the head-related transfer function, whereby the shapes and locations of the head, ears, and other anatomical features transform how the sound source originating from a particular location is finally perceived by a human listener. A simplified understanding of this phenomenon, known as the duplex theory~\cite{Rayleigh1907OnDirection}, can be thought of in terms of the interaural time differences of arrival, interaural level differences, and interaural correlation between the acoustic signals received at each ear.
The duplex theory, in particular, has been exploited to achieve significant data compression in various perceptual audio codecs \cite{InternationalOrganizationforStandardization1998ISO/IECAudio,InternationalOrganizationforStandardization2006ISO/IECAAC,Xiph.OrgFoundation2020VorbisSpecification,Valin2012RFCCodec}. These techniques, however, are known to be capable of producing audible spatial artifacts, especially at lower bitrates \citep{Brandenburg1999MP3Explained, Herre2004FromStandardization, Rumsey2005OnQuality, Delgado2019ObjectiveMaps, Delgado2019InvestigationsQuality}.

Spatial audio quality is a highly multifaceted concept \cite{Lindau2014ASAQI}, with some aspects that are inherently perceptual, thus requiring subjective listening tests. However, objective aspects of the spatial quality, particularly geometric ones, continue to be an underexplored avenue for objective analytical methods to obtain quantitative parametrization of the spatial impairment, even if these do not immediately correlate to perceptual constructs. To the best of our knowledge, few experimental studies have specifically investigated objective measurements of these spatial artifacts.
Unsurprisingly, equally few objective metrics have been designed specifically for spatial evaluation of multichannel audio. Most practical evaluation of spatial quality continues to rely on time-consuming and labor-intensive listening tests, often requiring expert listeners \cite{Lindau2014ASAQI}.
However, a number of spectral and spatial features were proposed for training a simple model to predict perceptual ratings \cite{George2006FeatureFidelity,George2006InitialRecordings, Choi2007PredictionSystems,Choi2008ObjectiveSystems, Delgado2019ObjectiveMaps}. 
More recently, a similar deep neural metric was proposed in \cite{Manocha2022SAQAM:Metric} for binaural audio. All of these data-dependent approaches, however, are not usable outside the channel configuration in which the prediction models were trained on. The generalizability of the models is also often called into question when applying the predictor to unseen data with significant domain shifts. AMBIQUAL, a metric designed for ambisonic signals derived from structural similarities of the time-frequency representations \citep{Narbutt2018AMBIQUALAudio, app10093188}, does not suffer from data dependency but was designed only for ambisonic data.

Despite limited literature specific to spatial evaluation, findings from audio quality evaluation in other subdomains can be adapted for spatial audio evaluation. In source separation, the BSS Eval toolbox~\citep{Vincent2006PerformanceSeparation, Vincent2007FirstResults, Vincent2012TheChallenges} has been used widely to measure various aspects of signal degradation due to the separation algorithms. To account for filtering error, the toolbox computes a 512-tap least-square multichannel projection filter $\mathbf{h}$ of reference signal $\mathbf{s}$ onto the test signal $\hat{\mathbf{s}}$. The resulting error signal $\mathbf{e}_\text{proj} = \mathbf{s} \ast \mathbf{h} - \hat{\mathbf{s}}$ has often been referred to as the ``spatial error'', despite containing all filtering errors accountable within 512 taps regardless of their spatial relevance. This issue, in particular, has led to the limited utility of the ``source image to spatial distortion ratio'' (ISR), relative to other metrics in BSS Eval. 

By constraining the projection filter, however, it is possible to exclusively account for spatial distortions, in particular those accounted for by the duplex theory, due to their frequency independence. Spatial information originating from the room acoustics or head-related transfer functions, however, cannot be easily distinguished \textit{a priori} from other frequency-dependent distortion. In this work, we propose a decomposition technique\footnote{Implementation and additional derivations are available from \href{https://github.com/karnwatcharasupat/spauq}{github.com/karnwatcharasupat/spauq}. Last accessed: \lastaccessed.} that distinguishes a subset of spatial distortions from other filtering distortion, allowing the energy ratio between a clean signal and the corresponding spatial error signal to be explicitly computed. The technique was designed such that it is task-agnostic and can be used either independently for, e.g., codec evaluation, or in conjunction with existing ratio metrics for evaluation of source separation systems.

\section{Proposed Method} \label{sec:method}


In a multichannel audio setting, the filtering error itself can be loosely decomposed into one concerning spatial distortion, such as interchannel time differences (ITD) and interchannel level differences (ILD), and another concerning frequency response distortion, such as changes in equalization (EQ) or timbre. Admittedly, any spatial effect with a frequency-dependent response, such as those due to room reverberation or the pinna filtering effects, cannot be fully distinguished from EQ distortion. As such, only  filtering operations related to the duplex theory will be considered in this work.


Changes to the ITD can be modeled by relative changes in delays over the channels, while changes to the ILD can be modeled by relative changes in the gain of each channel as well as leakages into other channels. Changes to the interchannel correlation are inherently computable from the ITD and ILD and thus do not need explicit modeling. For a signal $\mathbf{s}$ with $C$ channels and $N$ samples, the projected signal $\tilde{\mathbf{s}}$ can thus be modeled by
\begin{equation}
   \textstyle  \tilde{s}_c[n] = \sum_{d=1}^{C} A_{cd}s_d[n - \tau_{cd}], \label{eq:decomp}
\end{equation} 
for valid zero-indexed samples $n\in\mathfrak{N}_c$ for each one-indexed channel $c\in\mathfrak{C}\coloneqq\llbracket1,C\rrbracket$, where $\mathfrak{N}_{c}=\bigcap_{d\in\mathfrak{C}}\mathfrak{N}_{c,d}$, and $\mathfrak{N}_{c,d} = \llbracket \max(0, \tau_{cd}), N + \min(0, \tau_{cd})\rrparenthesis$. We also denote $\mathfrak{N}=\bigcap_{c\in\mathfrak{C}}\mathfrak{N}_{c}$.

Since we are interested in projecting the \textit{reference} signal as close to the \textit{estimated} signal as possible, this results in the least-square optimization objective 
\begin{align}
    \textstyle \min_{\mathbf{A}, \mathbf{T}} \sum_{n\in \mathfrak{N}} |\hat{\mathbf{s}}_c[n]-\tilde{\mathbf{s}}_c[n]|^2
\end{align}
where $\mathbf{A},\mathbf{T}\in\mathbb{R}^{C\times C}$, 
$(\mathbf{T})_{cd}=\tau_{cd}$.
Note that the resulting filter is a multichannel filter with at most one non--zero tap in each filter channel. This can be considered as a one-hot special case of the BSS Eval filter, but with a different limit on the maximum number of taps.

\subsection{Solving for Optimal Parameters}


Denote a generalized correlation operator by
\begin{align}
   \textstyle  \mathcal{R}_{\mathfrak{I}}(u, v)[\nu, \eta] &= \textstyle \sum_{n\in\mathfrak{I}}u[n - \nu]v[n-\eta],
\end{align}
where
$r_{cd}[\eta] = \mathcal{R}_{\mathfrak{M}_\eta}(s_c,s_d)[0, \eta]$, $\hat{r}_{cd} = \mathcal{R}_{\mathfrak{M}_\eta}(\hat{s}_c,\hat{s}_d)[0, \eta]$, $\check{r}_{cd} = \mathcal{R}_{\mathfrak{M}_\eta}(\hat{s}_c, s_d)[0, \eta]$, $\breve{r}_{cd} = \mathcal{R}_{\mathfrak{M}_\eta}(h\ast\hat{s}_c, h\ast s_d)[0, \eta]$, $h$ is an optional lowpass filter,
and $\mathfrak{M}_\eta = \llbracket{\max(0,\eta)}, N+\min(0,\eta)\rrparenthesis$. Solving for $\mathbf{T}$ directly remains an open problem due to multiple local minima and non-monotonic gradients. 
As such, we constrain $\mathbf{T}$ to the integral space $\mathbb{Z}^{C\times C}$ and used interchannel correlation to assign
\begin{align}
    \tau_{cd} = \underset{-K \le \kappa \le K}{\operatorname{arg\ max}} \left| \underset{f \in \mathfrak{F}}{\mathrm{IDFT}}\left\{
        \hat{S}_c[f] \cdot S^\ast_d[f] \cdot |H[f]|^2\right\}[\kappa] \right|,
\end{align}
where $K\in\mathbb{Z}^+$ is the search limit, $\mathfrak{F}$ is the set of frequency indices, and $H$, $\hat{S}_c$ and $S_d$ are DFTs of $h$, $\hat{s}_c$ and $s_d$ computed with appropriate zero-padding. We defaulted $K$ to the discrete-time equivalent of \SI{50}{\milli\second}, which is well above the human spatialization TDOA limit. In other words, each input channel of the reference signal is shifted so that it is maximally correlated to the target channel of the test signal or its inversion. 



At an optimal $\mathbf{T}$, the optimal value for each row of $\mathbf{A}$ can be found by solving the matrix equation
$\mathbf{A}_{c,:}\mathbf{R}^c = \check{\mathbf{R}}_{c, :}$,
where $(\mathbf{R}^c)_{bd} = \mathcal{R}_{\mathfrak{N}}(s_b, s_d)[\tau_{cb}, \tau_{cd}]$,
and $(\check{\mathbf{R}})_{cd} = \check{r}_{cd}[\tau_{cd}]$. Since $\mathbf{R}^c$ is symmetric, we simply 
use matrix inversion when it is numerically stable.
In practice, however, some channels of the reference and/or test signals can be (nearly) silent, leading to numerical instability. To address this, we first set $\mathfrak{K} \coloneqq \llbracket1,C\rrbracket$, $\mathfrak{D} \coloneqq \llbracket1,C\rrbracket$, and a threshold $\epsilon \in \mathbb{R}^+$.
When $\sum_{n}\hat{s}^2_c[n] < \epsilon$, we set $\mathbf{A}_{c,:} \gets \mathbf{0}$ and $\mathfrak{K}\gets \mathfrak{K} \backslash \{c\}$. 
When $\sum_{n}s^2_d[n] < \epsilon$, we set $\mathbf{A}_{:,d} \gets \mathbf{0}$ and $\mathfrak{D}\gets \mathfrak{D} \backslash \{d\}$. We then solve 
\begin{align}
    \mathbf{A}_{c,\mathfrak{D}} &= \check{\mathbf{R}}_{c, \mathfrak{D}}\left(\mathbf{R}^{c}_{\mathfrak{D},\mathfrak{D}}\right)^{-1}, \quad \forall c \in \mathfrak{K}.
\end{align}

\subsection{Energy-Ratio Metrics}

Once the optimal projection $\tilde{\mathbf{s}}$ is found, the spatial error signal can be computed using $\mathbf{e}_\text{spat} = \tilde{\mathbf{s}} - \mathbf{s},$
while any other residual error can be computed by treating $\tilde{\mathbf{s}}$ as the `new' reference, i.e., $\mathbf{e}_\text{resid} = \hat{\mathbf{s}} - \tilde{\mathbf{s}}$. Thus, the total error between the reference and the test signal can be written as $\mathbf{e}_\text{total} = \hat{\mathbf{s}} - \mathbf{s} = \mathbf{e}_\text{spat} +  \mathbf{e}_\text{resid}.$
Using the decompositions above, two metrics naturally arise, which we refer to as the Signal to Spatial Distortion Ratio (SSR),
\begin{align}
    \text{SSR}(\hat{\mathbf{s}}; \mathbf{s})
        &= 10 \log_{10} \left(\|\mathbf{s}\|^2/{\|\mathbf{e}_\text{spat}\|^2}\right),
\end{align}
and the Signal to Residual Distortion Ratio (SRR),
\begin{align}
    \text{SRR}(\hat{\mathbf{s}}; \mathbf{s})
        &= 10 \log_{10} \left(\|\tilde{\mathbf{s}}\|^2/{\|\mathbf{e}_\text{resid}\|^2}\right).
\end{align}
The SSR itself can be considered as a replacement for the ISR, considering only components of the error signals with spatial importance as errors. The SRR effectively acts as the non-spatial SNR, only considering non-spatial errors such as interference, timbral distortion, and additive artifacts.


\subsection{Framewise Computation}

Since the proposed decomposition is relatively easy to compute, it can be implemented in a frame-wise manner. This is particularly helpful in the case of time-variant signals such as music, speech, and environmental sound where the signal content can drastically change over a time period. This means that most audio processing algorithms may also process the signal in a time-variant manner, leading to time-varying spatial distortion which in turn requires time-varying decomposition. Following BSS Eval, we defaulted to a window of \unit{2}{s} with 50\% overlap in our implementation.

\begin{figure*}[t]
\centering
    \centering
    \subfloat[]{\includegraphics[height=1.3in]{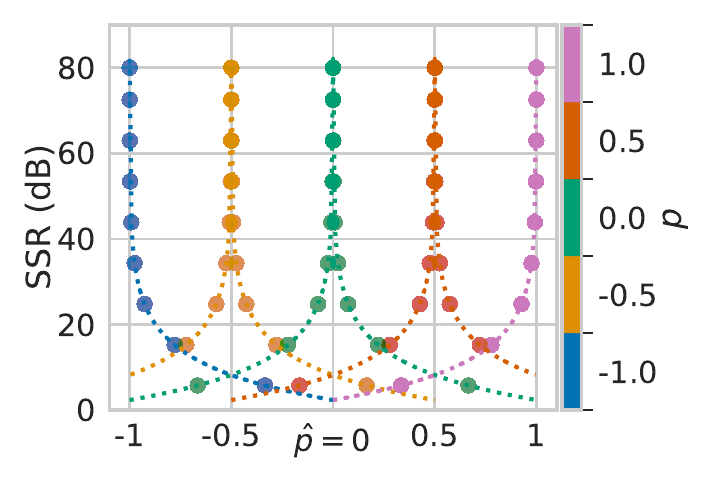}\vspace{-2ex}}
    \subfloat[]{\includegraphics[height=1.3in]{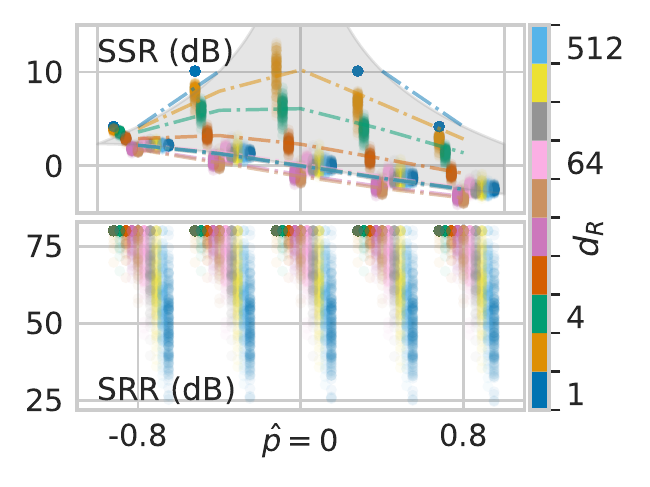}\vspace{-2ex}}
    \subfloat[]{\includegraphics[height=1.3in]{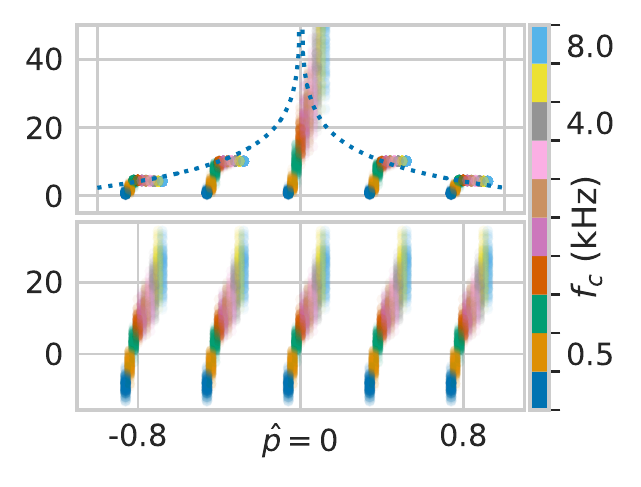}\vspace{-2ex}}
    \subfloat[]{\includegraphics[height=1.3in]{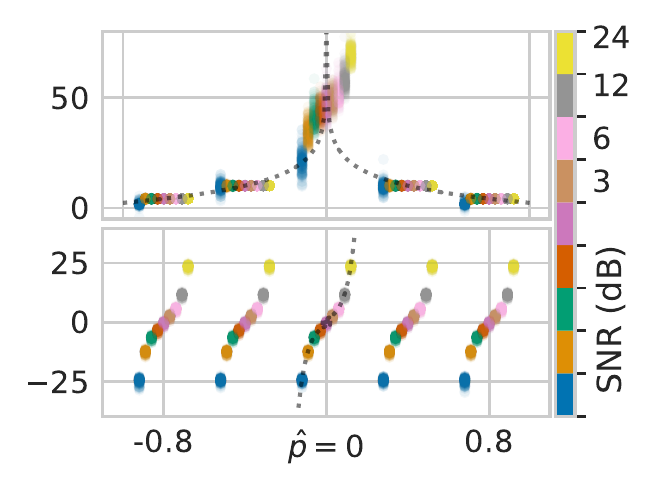}\vspace{-2ex}}
    \vspace{-8pt}
    \caption{SSR and SRR of the test signals w.r.t. its panning and (a) reference signal panning; (b) right-channel delay; (c) cutoff frequency; (d) SNR. Circular markers are experimental values with the horizontal offsets for readability. (b) In the SSR plot, each dashdotted line connects the median values within a delay parameter; the gray area represents the theoretical range of the SSR. (c \& d) Dotted lines are theoretical values.\label{fig:comp}
    }
\end{figure*}

\section{Signal Degradation Tests}

To evaluate the robustness of the proposed decomposition and thus the proposed metrics, common signal degradations and spatial distortions are evaluated on a subset of the TIMIT Acoustic-Phonetic Continuous Speech Corpus \cite{garofolo1993timit}. 
For the purpose of this robustness check
we test various audio degradation on recorded utterances of the sentence \texttt{SA1}, as uttered by 168 different participants with various dialectical variants of American English; \texttt{SA1} is chosen as it was designed to expose diverse variations in English phoneme pronunciation.
The TIMIT Corpus provides single-channel 16-bit PCM audio signals sampled at \SI{16}{\kilo\hertz}. In order to simulate known spatialization settings, each mono signal is spatialized to a stereo setup via the constant-power pan law
$g_\text{L} = \cos\left(\frac{\pi}{4}(p+1)\right)$, $
    g_\text{R} = \sin\left(\frac{\pi}{4}(p+1)\right)$, with~$p\in [-1, 1]$. 

\begin{figure}[tb]
    \centering
    \includegraphics[width=0.9\columnwidth]{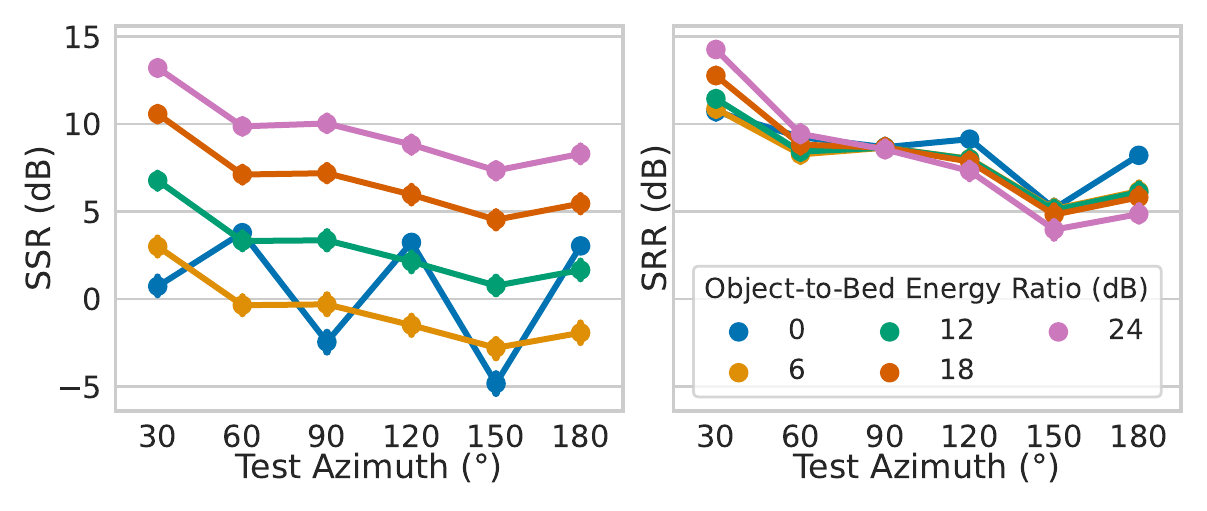}
    \vspace{-10pt}
    \caption{SSR and SRR of the test signals w.r.t. its azimuthal locations and the object-to-bed energy ratio. Each circular marker represents the mean over frames and bed-object pairs; each vertical line is the \SI{95}{\percent} confidence interval of the mean.}
    \label{fig:starss}
\end{figure}

\subsection{Panning Error}

The most basic test is to investigate the relationship between the metrics in the case where there is only spatial error and no other type of degradation. This first test is simulated by considering magnitude-only stereo panning errors between the test signals and the reference signals. As theoretically expected, all computed SRR values on the \texttt{SA1} signals of the TIMIT Corpus test set were all numerically positive infinities. The SSR results are shown in \Cref{fig:comp} (a) and are consistent with the theoretical values with very small variances.

\subsection{Delay Error}

The next test investigates the channel-wise delay error. The reference signal for this experiment is center-panned ($p=0$) with no delay applied. The test signals are panned at various errors, with an additional delay of $\hat{d}_\text{R}$ samples applied only on the right channel. The results of the delay test are shown in \Cref{fig:comp} (b). 
Since the theoretical SSR is dependent on the autocorrelation function (ACF) of the signal, the theoretical range for a zero-mean signal was shown as a shaded region. The theoretical value of SRR is positive infinity.

In general, most experimental values lie within or close to the expected range. Within each pan parameter, the SSR follows roughly the shape of the ACF of a speech signal with minima roughly at the delay where the shifted speech signal would be at maximally negative correlation with the unshifted version of itself. 
With the exception of $\hat{p}=-1$, where only the left channel of the estimate is present, the distributions of the SRR are nearly identical across all pan values. At lower delays, more SRR values are at concentrated the expected positive infinity (capped at \SI{80}{dB} for numerical stability). As the delays increase, the SRR values tend to decrease, but remain at relatively high values above \SI{25}{dB}. Upon inspection of the decomposition, the channel-wise shift values have been estimated correctly for all test signals while the channel-wise gain values are often only \textit{approximating} the ideal values, with the deviation increasing with $\hat{d}_\text{R}$. The slightly imperfect projection thus results in the observed spread and deviation in the SRR values from the theoretical values.

\subsection{Filtering Error}

Many audio processing algorithms can cause a loss of bandwidth \cite{Brandenburg1999MP3Explained, Schaffer2022MusicModeling}. To test the ability of the proposed method to distinguish other filtering errors from spatially relevant ones, the estimates were forward-backward filtered with a 128-tap low pass filter computed using the Remez exchange algorithm at various cutoff frequencies $f_\text{c}$ with a transition band of one third-octave. The reference signal for this experiment is center-panned ($p=0$) with no delay applied. 
The results of this test are shown in  \Cref{fig:comp} (c). As expected, the SRR increases monotonically as the cutoff frequency increases. At high cutoff frequencies, the SSR is close to the theoretical value. As more of the signal content is lost with decreasing cutoff frequency, the SSR also decreases, with a large deviation below \SI{1}{\kilo\hertz}. This is somewhat expected given that spatial decomposition will only be valid if most of the signal content is present.

\subsection{Additive Noise}

Another common audio degradation is the addition of noise or other uncorrelated artifacts. In the presence of an uncorrelated additive noise, the SRR is theoretically the overall SNR itself. The computed metrics after adding random Gaussian noise at various SNRs are shown in  \Cref{fig:comp} (d). The reference signal for this experiment is center-panned ($p=0$) with no delay applied. At $\hat{p}\ne0$, Where the theoretical values are finite, the SSR generally follows the theoretical values with small spreads except for the most noisy case where the SNR is \SI{-24}{dB}, demonstrating that the decomposition is generally robust to noise. The experimental SRR values largely follow the SNRs themselves with small spreads, demonstrating that the residual errors consist almost entirely of non-spatial errors.

\subsection{Multichannel Test with Real Sound Scenes as Beds}

We additionally tested the proposed method on a 5-channel spatialized audio with bed signals drawn from the test set of STARSS22 \cite{Politis2022STARSS22:Events} and object signals drawn from TIMIT. Non-stationary multichannel signals in STARSS22 were recorded in real-world spaces with Eigenmike em32. All signals were first normalized to \SI{-24}{LUFS} and upsampled to \SI{48}{\kilo\hertz} for compatibility with SPARTA. TIMIT signals were first spatialized to azimuths $\{\SI[mode=text]{0}{\degree}, \SI[mode=text]{30}{\degree}, \dots, \SI[mode=text]{180}{\degree}\}$ and elevation \SI{30}{\degree} in first-order ambisonics (FOA) format. All signals are then decoded to a typical 5-channel setup (\SI{0}{\degree}, \textpm\SI{30}{\degree}, \textpm\SI{110}{\degree}) using SPARTA\footnote{\href{https://leomccormack.github.io/sparta-site/}{leomccormack.github.io/sparta-site}. Last accessed: \lastaccessed.}. 
In each evaluation pair, a superposition of a bed signal and a object signal spatialized at \SI{0}{\degree} azimuth at a particular object-to-bed energy ratio (OBER) is used as the reference signal, while that of the same bed signal and the same object signal spatialized at another azimuthal position are used as the test signals. 

Experimental results are shown in \Cref{fig:starss}. Across all aggregation points, the confidence intervals are relatively small, indicating that the proposed methods return relatively stable results across time and bed-object pairs, despite the beds containing many different moving sounds and being recorded across five different acoustic environments. In terms of SSR, it can be seen that at all OBERs except \SI{0}{\decibel}, the SSR trends over the test azimuths are very similar, capturing the expected spatial degradation given the speaker setup. In terms of SRR, it can be seen that the proposed method is rather robust to the OBERs and returns similar results at each test azimuth position. The sudden drop in SSR at \SI{150}{\degree} can likely be attributed to the poor speaker coverage at that angle.

\section{Benchmarks}

As a benchmark for the proposed metrics, we apply perceptual audio compression and music source separation algorithms on the \mbox{MUSDB18-HQ} dataset \cite{musdb18}, which provide 50 uncompressed stereo music signals sampled at \SI{44.1}{\kilo\hertz}. Audio compression and music source separation are specifically chosen as test cases as these are nonlinear and waveform-systems that are known to introduce both spatial and non-spatial artifacts on music signals.

\subsection{Codec}

We apply AAC (FAAC 1.30; FAAD2 2.10.0-2), and Opus (libopus 1.3.1) to the test set of \mbox{MUSDB18-HQ} to investigate their impact on SSR and SRR at bitrates from \num{32} to \SI{320}{Kbps}. For AAC, the changes in SSR and SRR at each bitrate compared to no joint encoding are shown in \Cref{fig:daac}. Additional plots are provided in the Supplementary Materials\footnotemark[1]. For both AAC and Opus, both SSR and SRR increased as the bitrate increased. The trend in SSR is consistent with the literature on Opus \cite{Narbutt2017StreamingAudio, Rudzki2019AuditoryScenes} where localization errors increase with decreasing bitrate. In AAC, both mid/side stereo (MS) and intensity stereo (IS) generally performed worse in SSR than no joint coding across most ABRs, except for the MS mode at very low ABRs of \SI{32}{Kbps} and \SI{40}{Kbps}. In particular, IS also consistently performed worse than MS up to an ABR of \SI{192}{Kbps}. This is consistent with the knowledge that IS can cause severe spatial artifacts, especially for low-frequency content with decorrelated spatial images \cite{Herre1994IntensityCoding, Brandenburg1999MP3Explained, Herre2004FromStandardization}. In terms of SRR, which effectively measures the non-spatial fidelity of the codec, MS performed better than no joint coding up to about \SI{112}{Kbps} while intensity stereo only performed better than no joint coding below \SI{64}{Kbps}. It was expected that no joint coding performed better than joint coding from about \SI{128}{Kbps} onwards since joint coding can introduce unnecessary information loss at these bitrates \cite{Herre2004FromStandardization}.

\begin{figure}[t]  
    \centering
    \includegraphics[width=0.9\columnwidth]{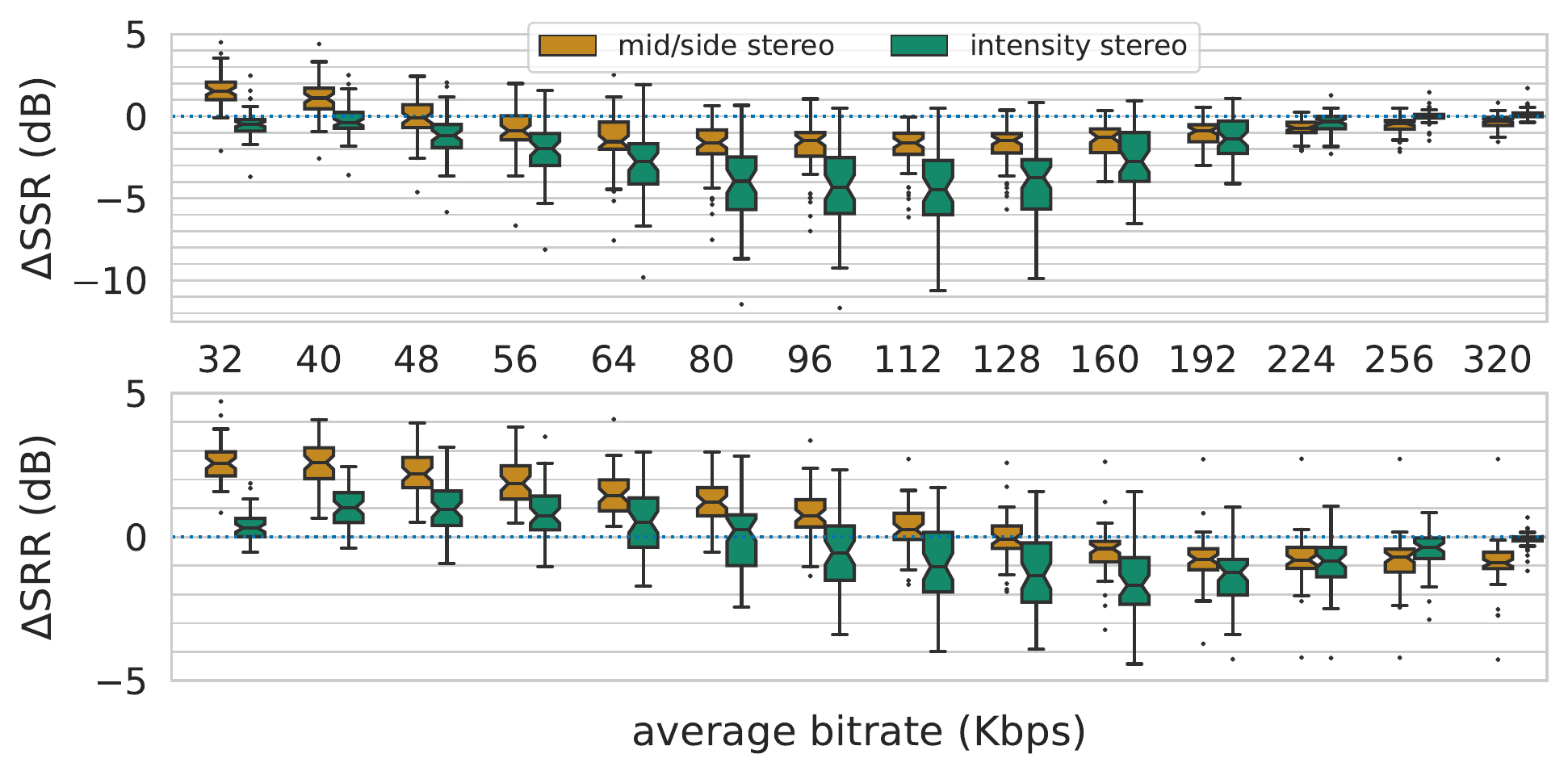}
    \vspace{-10pt}
    \caption{Change in SSR and SRR of the test signals compressed by AAC, relative to the operating mode without joint encoding, by operating mode and average bitrates.}
    \label{fig:daac}
\end{figure}

\begin{figure}[t]
    \centering
    \includegraphics[width=0.9\columnwidth]{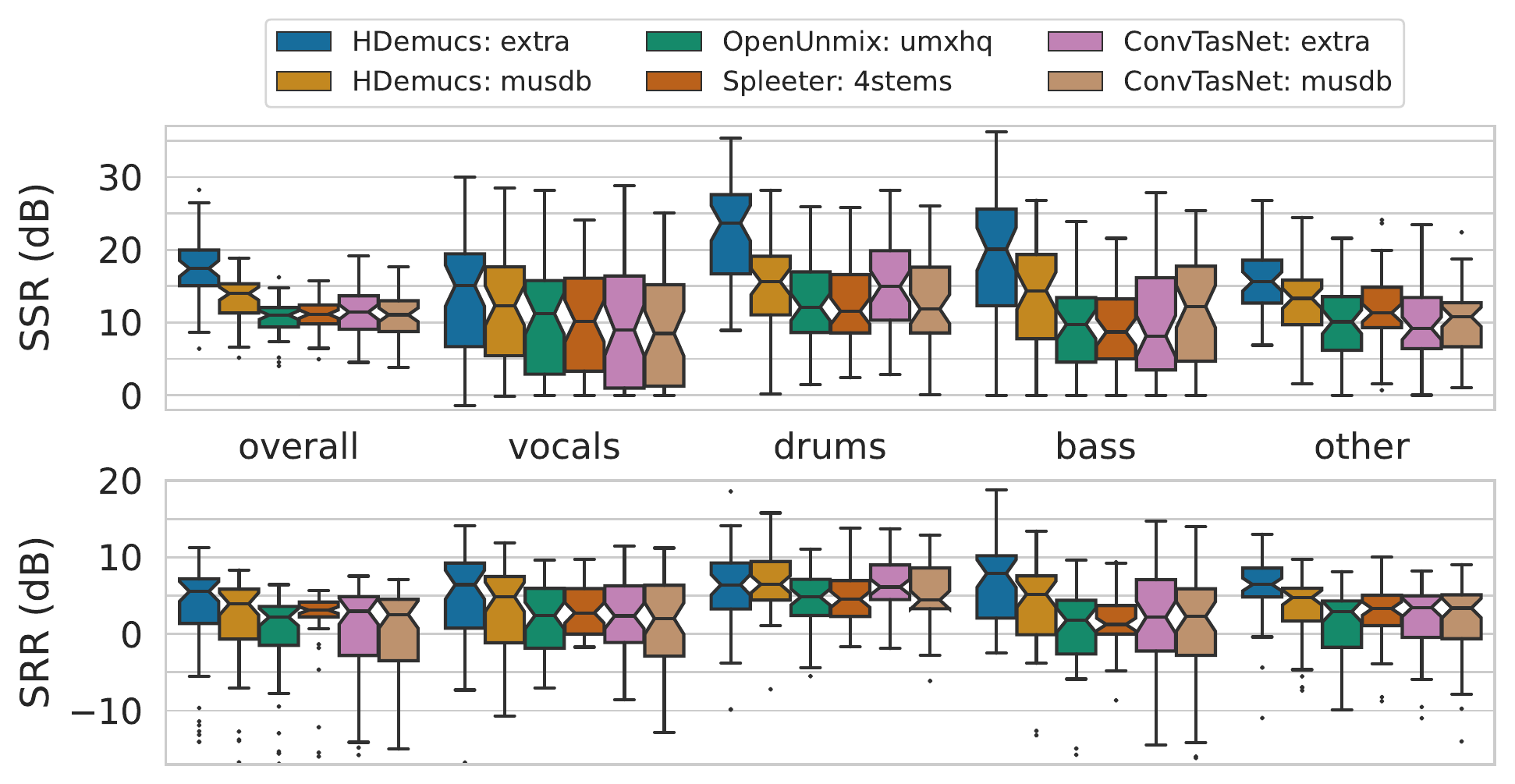}
    \vspace{-10pt}
    \caption{Evaluation results on the \mbox{MUSDB18-HQ} test set.}
    \label{fig:musdb}
\end{figure}

\subsection{Music Source Separation}

To benchmark the proposed metrics on the music source separation task, we apply Hybrid Demucs (v3) \cite{Defossez2021HybridSeparation}, 
ConvTasNet \cite{Luo2019Conv-TasNet:Separation}, 
OpenUnmix \cite{Stoter2019Open-UnmixSeparation}, and
Spleeter \cite{Hennequin2020Spleeter:Models} on the test set of \mbox{MUSDB18-HQ}. OpenUnmix and Spleeter perform separation in the time-frequency domain using real-valued channel-wise masks on the complex-valued short-time Fourier transform (STFT) spectrogram. The results reported here for both OpenUnmix and Spleeter are computed without the multichannel Wiener filter (MWF) postprocessing. ConvTasNet similarly performs real-valued mask-based separation but on a learnable real-valued basis transform. Hybrid Demucs is the only model tested here that does not utilize masking, instead modifying the time-domain signal and the STFT representation directly in its time and time-frequency branches, respectively. Note that OpenUnmix and Spleeter were optimized in the time-frequency domain without considering phase information, while ConvTasNet and Hybrid Demucs were optimized in the time domain. 

The performance of the models is shown in \Cref{fig:musdb}. The suffix behind the model name refers to the pre-trained weight variants provided by the model developers\footnote{The training data of HDemucs:extra also contains the test set of \mbox{MUSDB18-HQ} thus may not provide a fair comparison to the rest of the models which has not seen the test set in its training data.}. The SRR values, which effectively act as a non-spatial counterpart of the SDRs, are consistent with the SDRs reported in the literature. In terms of SSR, Demucs performs the best among the tested models, while other models perform approximately on par with one another. We surmise that the superior performance of Demucs may be due to either or a combination of 
\begin{inparaenum}[(i)]
    \item its non-masking nature, 
    \item the use of a direct time-domain processing branch, and/or 
    \item direct optimization in the time domain. 
\end{inparaenum}

\section{Conclusion}

In this work, we proposed a novel spatial evaluation method using a filter decomposition technique based on the duplex theory of spatial hearing. Tests on common signal degradation demonstrated robust performance. The proposed method is benchmarked on audio compression algorithms and music source separation systems, showing results consistent with expectations and literature. An open-source Python implementation of the proposed method is provided.

\FloatBarrier

\section{Acknowledgement}
The authors would like to thank Chih-Wei Wu and Phillip A. Williams for their assistance with the project.

\section{References}

\renewcommand{\bibsection}{}

{\bibliographystyle{IEEEbiba}
\fontsize{8}{9.6}\selectfont
\setlength{\bibsep}{1pt plus 3pt}
\bibliography{references}}

\end{document}


\newcommand{\papertitle}{\large \textbf{Supplementary: Quantifying Spatial Audio Quality Impairment}}

\title{\papertitle}

\author{Karn N. Watcharasupat and Alexander Lerch}

\markboth{Submitted to ICASSP 2024}
{Watcharasupat and Lerch: \papertitle{}}
\maketitle

\newtheorem{thm}{Theorem}[section]
\newtheorem{lem}[thm]{Lemma}
\newtheorem*{deriv}{Derivation}
\newtheorem{prop}[thm]{Proposition}
\newtheorem{cor}{Corollary}
\newtheorem{defn}{Definition}[section]
\newtheorem{conj}{Conjecture}[section]
\newtheorem{exmp}{Example}[section]
\newtheorem{rem}{Remark}

\renewcommand{\thesection}{\Alph{section}}
\renewcommand{\theequation}{\thesection.\arabic{equation}}
\renewcommand{\thefigure}{A\arabic{figure}}
\section{Derivation of the Optimal Gain Matrix}
Denote the objective function to be minimized by
\begin{align}
    \mathcal{L}(\mathbf{A}, \mathbf{T}) 
    &=\sum_{n}\|\tilde{\mathbf{s}}[n]-\hat{\mathbf{s}}[n]\|^2\\
    &= \sum_{n}(
        \tilde{\mathbf{s}}^\T[n]\tilde{\mathbf{s}}[n]
        - 2 \hat{\mathbf{s}}^\T[n]\tilde{\mathbf{s}}[n]
        + \hat{\mathbf{s}}^\T[n]\hat{\mathbf{s}}[n]
        ).
\end{align}
Taking the derivative with respect to $\mathbf{A}$ gives
\begin{align}
    \dfrac{\de \mathcal{L}}{\de A_{ij}}
    &= 2\sum_{n}(
        \tilde{s}_i[n]
        - \hat{s}_i[n])s_j[n - \tau_{ij}],
\end{align}
since
\begin{align}
     \dfrac{\de\tilde{s}_c[n]}{\de A_{ij}}
        &= \mathbb{I}[i=c] s_j[n - \tau_{ij}],
\end{align}
where $\mathbf{I}[\cdot]$ is the Iverson bracket. By setting the derivative to zero, we have
\begin{align}
    &\sum_n  \hat{s}_i[n] s_j[n - \tau_{ij}]
    =\sum_{n} \tilde{s}_i[n]s_j[n - \tau_{ij}]\\
    &\qquad\qquad=\sum_{d}A_{id} \sum_{n} s_d[n-\tau_{id}]s_j[n - \tau_{ij}]
\end{align}
which reduces to (5).

\section{Derivation of the Theoretical SSR}

For this section, let $\text{SSR} = 10\log_{10} u$. Denote the mono reference signal by $v$. Let $E_v = \sum_{n}v^2[n]$. Considering only the necessary parametrization for III.A and III.B, 
\begin{align}
    \hat{\mathbf{s}}[n] = \begin{bmatrix}\hat{g}_\text{L} v[n]\\ \hat{g}_\text{R} r[n - \hat{d}_\text{R}]\end{bmatrix}.
\end{align}
Since all spatial error are theoretically accountable by (1), ${\tilde{\mathbf{s}} \equiv \hat{\mathbf{s}}}$. The reference signal is given by 
\begin{align}
    \mathbf{s}[n] = \begin{bmatrix}g_\text{L}\\g_\text{R}\end{bmatrix} \cdot v[n],
\end{align}
and $E_\mathbf{s} =  \sum_{n}\|\mathbf{s}[n]\|^2 = E_v$ since $g^2_\text{L}+g^2_\text{R}=1$.

\subsection{Panning Error}
Where only panning error is present, $\hat{d}_\text{R}=0$. Therefore,
\begin{align}
    u &= \left(\sum_{n}\|\tilde{s}[n]-\mathbf{s}[n]\|^2\right)^{-1}\left(\sum_{n}\|\mathbf{s}[n]\|^2\right)\\
    u^{-1}E_v
    &= \sum_{n}\left\| \begin{bmatrix}\hat{g}_\text{L} - g_\text{L}\\\hat{g}_\text{R} - g_\text{R}\end{bmatrix} \cdot v[n]\right\|^2 \\
    &= \left[(\hat{g}_\text{L} - g_\text{L})^2+(\hat{g}_\text{R} - g_\text{R})^2\right] \cdot E_v \\
    u^{-1} &= 2 - 2\cos\left(\dfrac{\pi}{4}(\hat{p}-p)\right).
\end{align}

\subsection{Panning and Delay Error}
With $\hat{d}_\text{R} \ne 0$,
\begin{align}
    &u^{-1}E_v
    = \sum_{n}\left\| \begin{bmatrix}(\hat{g}_\text{L} - g_\text{L})v[n]\\\hat{g}_\text{R} v[n-\hat{d}_\text{R}]- g_\text{R}v[n]\end{bmatrix} \right\|^2 \\
    &\quad
    = (\hat{g}_\text{L} - g_\text{L})^2E_v + \sum_{n}\left| \hat{g}_\text{R} v[n-\hat{d}_\text{R}]- g_\text{R}v[n]\right|^2 \\
    &u^{-1} = (\hat{g}_\text{L} - g_\text{L})^2 + (\hat{g}_\text{R} - g_\text{R})^2 
    \nonumber\\&\qquad\qquad
    + \dfrac{2\hat{g}_\text{R}g_\text{R}}{E_v}\left[1 - \sum_n v[n-\hat{d}_\text{R}]v[n]\right]\\
    &u^{-1} = 2 - 2\cos\left(\dfrac{\pi}{4}(\hat{p}-p)\right) + 2\hat{g}_\text{R}g_\text{R}\left(1-\dfrac{\kappa_v[-\hat{d}_\text{R}]}{\kappa_v[0]}\right),
\end{align}
where $\kappa_v[\cdot]$ is the autocorrelation function of $v$. Since $|\kappa_v[d]| \le \kappa[0]$ for all $d$, we have
\begin{align}
    0 \le u^{-1} - 2 - 2\cos\left(\dfrac{\pi}{4}(\hat{p}-p)\right) \le 4\hat{g}_\text{R}g_\text{R}.
\end{align}

\section{Additional Experimental Results}
\begin{figure}[H]
    \centering
    \includegraphics[width=\columnwidth]{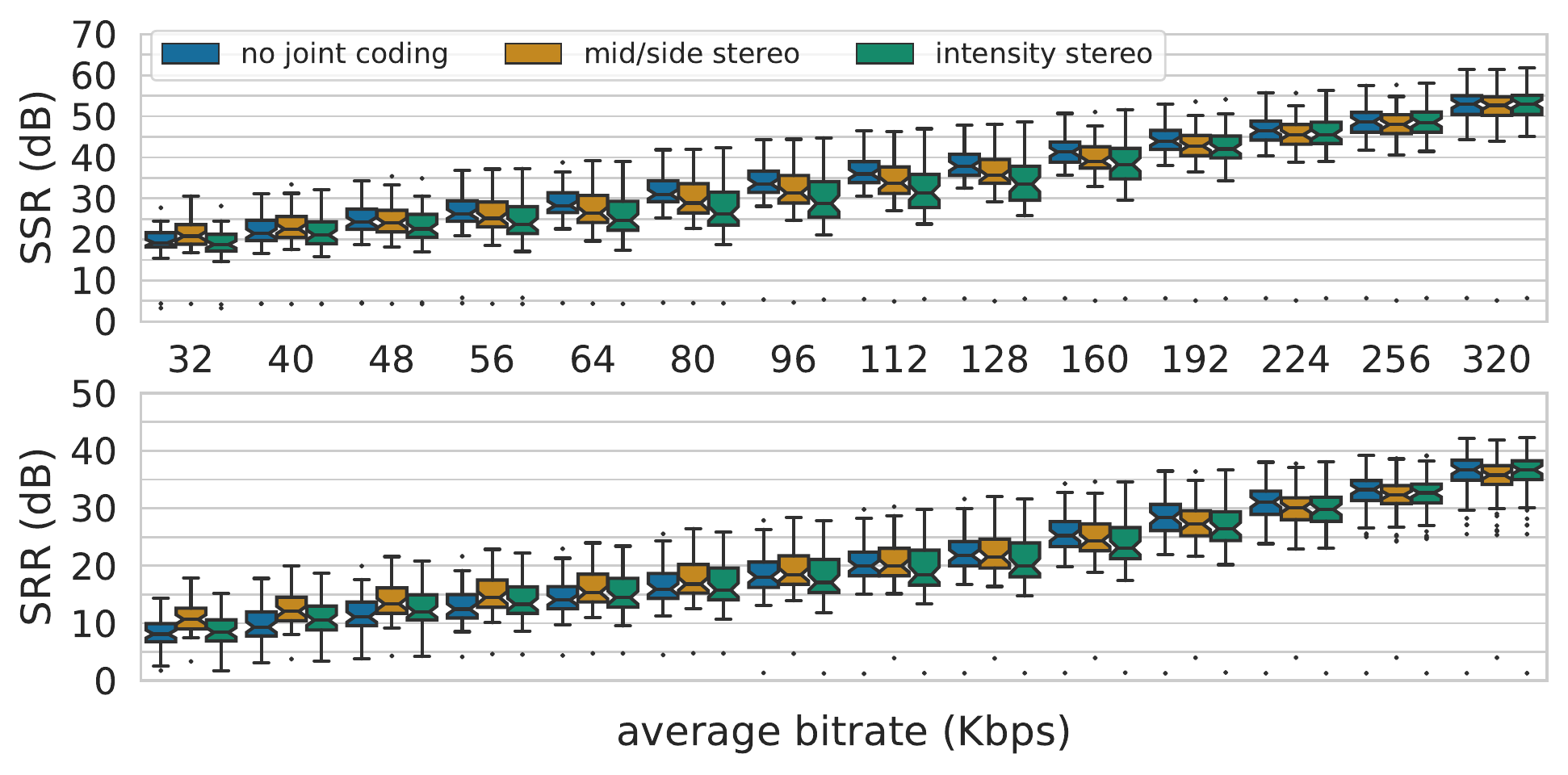}
    \caption{SSR and SRR of the test signals compressed by AAC, by operating mode and average bitrates.}
    \label{fig:aac}
\end{figure}

\begin{figure}[H]
    \centering
    \includegraphics[width=\columnwidth]{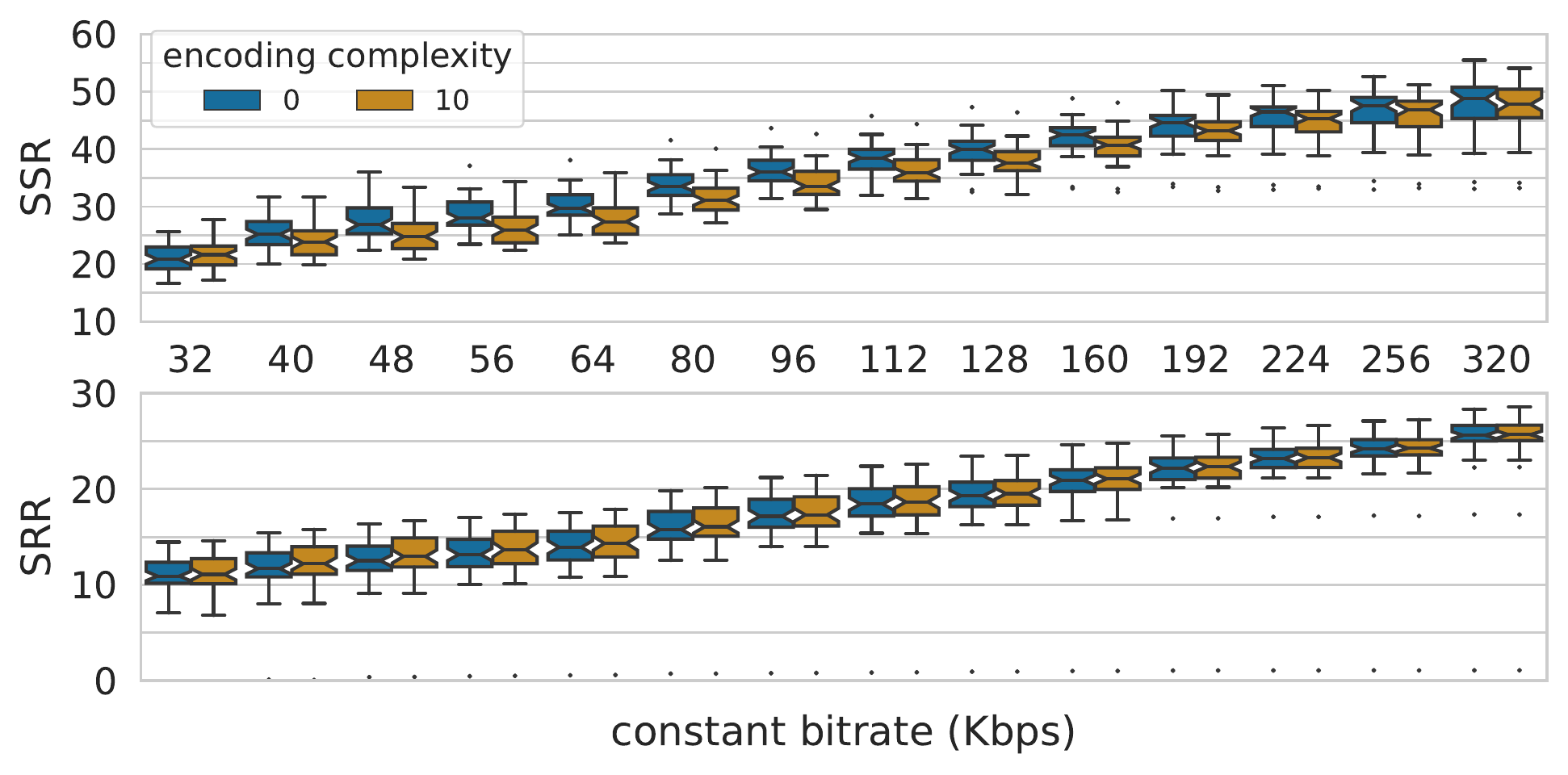}
    \caption{SSR and SRR of the test signals compressed by Opus, by operating mode and constant bitrates.}
    \label{fig:opus}
\end{figure}